\documentclass[twocolumn,showpacs,aps,preprintnumbers,amsmath,amssymb,floatfix]{revtex4}

\usepackage{graphicx}
\usepackage{dcolumn}
\usepackage{bm}

\newcommand{\met}{{\not \!\! E_T}}
\newcommand{\lsim}{\mathrel{\hbox{\rlap{\lower.55ex\hbox{$\sim$}}
\kern-.3em \raise.4ex \hbox{$<$}}}}

\begin{document}


\title{Precision sparticle spectroscopy in the inclusive
same-sign dilepton channel at LHC}

\author{Konstantin T.~Matchev${}^1$, Filip Moortgat${}^2$, 
Luc Pape${}^2$, and Myeonghun Park${}^1$}
\affiliation{${}^1$Physics Department, University of Florida, Gainesville, FL 32611, USA}
\affiliation{${}^2$ETHZ, Zurich, Switzerland}%

\date{27 July, 2010}

\begin{abstract}
The inclusive same-sign dilepton channel is already recognized as
a promising discovery signature for supersymmetry in the 
early days of the LHC. We point out that it can also be used for precision
{\em measurements} of sparticle masses after the initial discovery.
As an illustration, we consider the LM6 CMS study point in minimal supergravity,
where the same-sign leptons most often result from chargino decays to 
sneutrinos. We discuss three different techniques for determining
the chargino and sneutrino masses {\em in an inclusive manner}, 
i.e.~using {\em only} the two well measured lepton momenta, while
treating all other upstream objects in the event as a single
entity of total transverse momentum $\vec{P}_T$.
This approach takes full advantage of the large production rates 
of colored superpartners, but does not rely on the poorly measured 
hadronic jets, and avoids any jet combinatorics problems. We discuss the anticipated 
precision of our methods in the early LHC data.
\end{abstract}

\maketitle

A long standing problem in hadron collider phenomenology has been the determination of the 
absolute mass scale of new particles in events with missing energy.
The prototypical example of this sort is provided by any model of low-energy supersymmetry (SUSY)
with conserved $R$-parity, in which the lightest superpartner (LSP),
typically the lightest neutralino $\tilde \chi^0_1$, is a neutral, weakly interacting 
particle of a priori unknown mass \cite{Chung:2003fi}.
Astrophysics also adds credence to such scenarios, since
the LSP is a potential dark matter candidate, whose relic
abundance is typically in the right ballpark \cite{Bertone:2004pz}.
$R$-parity conservation guarantees that 
every event contains (at least) two invisible particles, 
whose energies and momenta are not measured, making the full 
reconstruction of such events a very challenging task.

Recently, several solutions to this problem at the Large Hadron Collider (LHC) 
have been proposed. Most of them rely on exclusive channels
\cite{masses}, where a sufficiently long decay 
chain can be properly identified. Unfortunately, this almost inevitably
requires the use of hadronic jets in some form in the analysis --
in most SUSY models, the main LHC signal is due 
to the strong production of colored superpartners, 
whose cascade decays to the neutral LSP necessarily involve hadronic jets. 
For many reasons, jets are notoriously difficult to deal with, 
especially in a hadron collider environment. 
Because of the high jet multiplicity in SUSY signal events, 
any jet-based analysis 
is bound to face a severe 
combinatorial problem 
and is unlikely to achieve any good precision. 
Thus it is imperative to have alternative methods 
which avoid the direct use of jets and instead rely only 
on the well measured momenta of any (isolated) leptons in the event.

\begin{figure}[t]
\includegraphics[width=7.5cm]{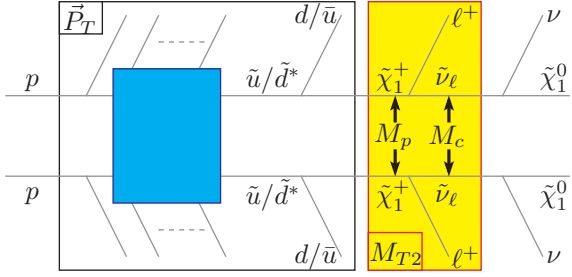}
\caption{\label{fig:event} The typical SUSY event topology producing two isolated
same-sign leptons at point LM6 (see text for details). 
The diagram for a pair of negatively charged leptons $\ell^-\ell^-$ is analogous.
}
\end{figure}

In this letter, we describe three such methods, which are free 
of the jet combinatorial problem.
For illustration, we shall use the standard example of 
$R$-parity conserving supersymmetry  with a $\tilde\chi^0_1$ LSP. 
Its collider signatures have been extensively studied, 
and typically involve jets, leptons and missing transverse 
energy \cite{Chung:2003fi}. Among those, the inclusive same-sign dilepton channel
has already been identified as a unique opportunity for an early 
SUSY discovery at the LHC \cite{Ball:2007zza,Aad:2009wy}. 
The two leptons of the same charge can be easily triggered on, and
provide a good handle for suppressing the SM background. 
In our analysis we use the LM6 CMS study point \cite{Ball:2007zza},
whose relevant mass spectrum is given in Table~\ref{tab:mass}. 
\begin{table}[t]
\caption{\label{tab:mass} Selected sparticle masses (in GeV) at point LM6. 
We list the average $\tilde q_L$ mass $M_{\tilde q_L}=\frac{1}{2}(M_{\tilde u_L}+M_{\tilde d_L})$.}
\begin{ruledtabular}
\begin{tabular}{cccccc}
$M_{\tilde g}$  & $M_{\tilde q_L}$ & $M_{\tilde \chi^+_1}$ &  $M_{\tilde \ell_L}$  & $M_{\tilde \nu_\ell}$    &  $M_{\tilde \chi^0_1}$ \\
\hline
939.8        & 862            &  305.3                  &  291.0                  & 275.7    &    158.1
\end{tabular}
\end{ruledtabular}
\end{table}
At point LM6, signal events with two isolated same-sign leptons typically 
arise 
from the SUSY event topology in Fig.~\ref{fig:event}.
Consider the inclusive production of same-sign charginos,
which decay leptonically 
as shown in the yellow-shaded box in the figure. The resulting sneutrino
($\tilde\nu_\ell$) could be the LSP itself, or, as in the case of LM6,
may further decay invisibly to a neutrino $\nu$ and the true LSP $\tilde\chi^0_1$. 
Such same-sign chargino pairs typically result from
squark decays, as indicated in Fig.~\ref{fig:event}.
In turn, the squarks 
may be produced directly through a $t$-channel gluino exchange, 
or indirectly in gluino decays. 
Note that the two same-sign leptons in Fig.~\ref{fig:event}
are accompanied by a number of upstream objects
(typically jets) which may originate from various sources, 
e.g.~initial state radiation, squark decays, or decays of even heavier 
particles up the decay chain. In order to stay clear of 
jet combinatorial issues, we shall adopt a fully 
inclusive approach to the same-sign dilepton signature,
by treating all the upstream objects 
within the black rectangular frame in Fig.~\ref{fig:event}
as a single entity of total transverse momentum $\vec{P}_T$.

Given this very general setup, we now pose the following question:
assuming that a SUSY discovery is made in the inclusive same-sign dilepton channel,
is it possible to measure the {\em individual} sparticle masses
$M_p$ and $M_c$ involved in  the leptonic decays of Fig.~\ref{fig:event},
using only the transverse momenta 
of the two leptons $\vec{p}_{\ell T}^{\, (1)}$ and $\vec{p}_{\ell T}^{\, (2)}$, and the 
{\em total} upstream transverse momentum $\vec{P}_T$?
Although it may appear that those three vectors 
do not provide a lot of information to go on, 
we shall show that this is possible.
We discuss three different approaches.

\begin{figure}[t]
\includegraphics[width=7.5cm]{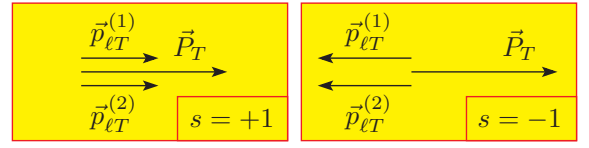}
\caption{\label{fig:mom} The two special momentum configurations 
defined in eqs.~(\ref{mom1},\ref{mom2}).}
\end{figure}

{\bf Method I.}
Let us concentrate directly on the observed lepton momenta $\vec{p}_{\ell T}^{\, (i)}$.
Consider the two collinear momentum configurations illustrated in 
Fig.~\ref{fig:mom} and defined as follows.
In each configuration, the lepton momenta are the same:
$\vec{p}_{\ell T}^{\, (1)}=\vec{p}_{\ell T}^{\, (2)}$; and 
then they can be either parallel or anti-parallel to the measured
upstream $\vec{P}_T$:
\begin{eqnarray}
s=+1
&\Rightarrow & 
\vec{p}_{\ell T}^{\,(1)}=\vec{p}_{\ell T}^{\,(2)} \uparrow\uparrow \vec{P}_T\, ;
\label{mom1}
 \\
s=-1 
&\Rightarrow & 
\vec{p}_{\ell T}^{\,(1)}=\vec{p}_{\ell T}^{\,(2)} \uparrow\downarrow \vec{P}_T .
\label{mom2}
\end{eqnarray}
In what follows we shall use the integer $s=+1$ ($s=-1$) to refer to the parallel
(anti-parallel) configuration:
$s\equiv \cos(\vec{p}_{\ell T}^{\, (1)},\vec{P}_T)=\cos(\vec{p}_{\ell T}^{\, (2)},\vec{P}_T)$.
Now let us measure the {\em maximum} lepton momentum in each configuration:
\begin{equation}
p_{\ell T}(sP_T) \equiv 
\max_{ {\vec{p}_{\ell T}^{\, (1)}=\vec{p}_{\ell T}^{\, (2)}}\, \wedge\ {\cos(\vec{p}_{\ell T}^{\, (1)},\vec{P}_T)=s}} \left\{ p_{\ell T}^{\, (i)} \right\}.
\label{eq:ptlepdef}
\end{equation}
Observe that {\em both} $p_{\ell T}(+P_T)$
and $p_{\ell T}(-P_T)$ can be directly measured from the lepton $p_T$ distributions.
For example, construct a 2D scatter plot $\{x,y\}$ of
\begin{equation}
x = \cos(\vec{p}_{\ell T}^{\, (1)}+ \vec{p}_{\ell T}^{\, (2)},\vec{P}_T), \quad
y = |\vec{p}_{\ell T}^{\, (1)} + \vec{p}_{\ell T}^{\, (2)}|,
\end{equation}
with the cut $|\vec{p}_{\ell T}^{\, (1)} - \vec{p}_{\ell T}^{\, (2)}|<\epsilon\, (\sim 0)$,
and take the limit
\begin{equation}
p_{\ell T}(sP_T) = \lim_{ x\to s} \left(\frac{y}{2}\right)\, .
\label{ptmeas}
\end{equation}

Armed with the two measurements $p_{\ell T}(+P_T)$
and $p_{\ell T}(-P_T)$, we can now directly solve 
for the masses $M_p$ and $M_c$. The formula for
$p_{\ell T}(sP_T)$ 
is
\begin{equation}
p_{\ell T}(sP_T) = \frac{ M_p^2-M_c^2}{4M_p^2}
\left( \sqrt{4M_p^2+(sP_T)^2}-sP_T\right)\, .
\label{eq:ptlep}
\end{equation}
Inverting (\ref{eq:ptlep}), we get
\begin{equation}
M_p 
= \frac{\sqrt{p_{\ell T}(-P_T) \, p_{\ell T}(+P_T)}}{p_{\ell T}(-P_T)-p_{\ell T}(+P_T)}
\, P_T\, ,
\label{Mptrue}
\end{equation}
thus fixing the {\em absolute} mass scale in the problem.
Once the parent mass $M_p$ is known, the child mass 
$M_c$ is 
\begin{equation}
M_c = M_p\sqrt{1-2\, \frac{p_{\ell T}(-P_T)-p_{\ell T}(+P_T)}{P_T}}\, .
\label{Mctrue}
\end{equation}
Thus we found 
the true sparticle masses $M_p$ and $M_c$ 
directly in terms of the measured lepton momenta $p_{\ell T}(\pm P_T)$  
and upstream momentum $P_T$. 
Note that the choice of the {\em value} for $P_T$ 
in eqs.~(\ref{Mptrue}) and (\ref{Mctrue}) is arbitrary, which
can be used to our advantage, e.g.~to
select the most populated $P_T$ bin, 
minimizing the statistical error.

{\bf Method II.} In our previous method, the lepton momenta $p_{\ell T}(\pm P_T)$ 
were measured directly from the data as implied by eq.~(\ref{ptmeas}).
Alternatively, we can obtain them indirectly from the endpoint of
the Cambridge $M_{T2}$ variable \cite{mt2}. To be more precise, 
we apply the ``subsystem'' $M_{T2}$ variable introduced in \cite{Burns:2008va}
to the purely leptonic subsystem in the yellow-shaded box of Fig.~\ref{fig:event}.
Following the generic notation of Ref.~\cite{Burns:2008va}, we denote
the input (test) mass of the sneutrino child as $\tilde M_c$.
The subsystem $M_{T2}$ variable is now defined as follows. 
First form the transverse mass $M_{T}$ for each (chargino) parent
\begin{equation}
M_{T}^{(i)} 
\equiv
\sqrt{\tilde M_c^2 + 2\left(
|\vec{p}_{\ell T}^{\, (i)}|\sqrt{\tilde M_c^2+|\vec{p}_{cT}^{\, (i)}|^2 }
- \vec{p}_{\ell T}^{\, (i)}\cdot\vec{p}_{cT}^{\, (i)} \right)}
\nonumber
\end{equation}
in terms of the assumed test mass $\tilde M_c$ and transverse momentum 
$\vec{p}_{c T}^{\, (i)}$ for each (sneutrino) child. Just like the traditional $M_{T2}$
\cite{mt2}, the leptonic subsystem $M_{T2}$ variable \cite{Burns:2008va}
is defined through a minimization procedure over all possible partitions 
of the unknown children momenta $\vec{p}_{cT}^{\, (k)}$, 
consistent with transverse momentum conservation 
$\sum_k (\vec{p}_{cT}^{\, (k)} + \vec{p}^{\,(k)}_{\ell T}) + \vec{P}_T =0$
\begin{equation}
M_{T2}(\tilde M_c, \vec{P}_T, \vec{p}_{\ell T}^{\, (i)}) 
\equiv 
\min_{}
\left\{\max\left\{M_{T}^{(1)},M_{T}^{(2)}\right\} \right\}\ .
\label{eq:mt2def}
\end{equation}
The $M_{T2}$ distribution has an upper kinematic endpoint 
\begin{equation}
M_{T2}^{max} (\tilde M_c, P_T) \equiv \max_{\rm all \ events} 
\left\{ M_{T2}(\tilde M_c, \vec{P}_T, \vec{p}_{\ell T}^{\, (i)}) \right\},
\label{eq:mt2maxdef}
\end{equation}
which can be experimentally measured and subsequently 
interpreted as the corresponding parent mass $\tilde M_p$ 
\begin{equation}
\tilde M_p(\tilde M_c, P_T) \equiv M_{T2}^{max} (\tilde M_c, P_T)\, ,
\label{eq:Mptilde}
\end{equation}
providing one functional relationship among $\tilde M_p$ and $\tilde M_c$, but 
leaving the individual masses still to be determined.

For us the importance of the $M_{T2}$ variable (\ref{eq:mt2def})
is that the momentum configurations in Fig.~\ref{fig:mom}
are precisely the ones which determine its endpoint $M_{T2}^{max}$. 
The complete analytical dependence of the $M_{T2}$ endpoint 
$\tilde M_p(\tilde M_c, P_T)$ on {\em both} of its arguments $\tilde M_c$ and $P_T$
is now known \cite{Burns:2008va}:
\begin{equation}
\tilde M_p (\tilde M_c, P_T)
=\left\{ 
\begin{array}{ll}
\tilde M_p (\tilde M_c, +P_T),  &   \textrm{if}~~ \tilde M_c\le M_c,   \\
\tilde M_p (\tilde M_c, -P_T),  &   \textrm{if}~~ \tilde M_c\ge M_c,
\end{array}
\right.
\label{eq:MT2maxLR}
\end{equation}
where 
\begin{eqnarray}
&& \tilde M_p( \tilde M_c, sP_T )
= \biggl\{ \biggl[ p_{\ell T}(sP_T)  \nonumber \\
&& +\sqrt{\left( p_{\ell T}(sP_T)+\frac{sP_T}{2}\right)^2+\tilde M_c^2} \, \biggr]^2 
-\frac{(sP_T)^2}{4} \biggr\}^\frac{1}{2}.~~~
\label{eq:Mp}
\end{eqnarray}
Thus we can alternatively obtain the sparticle masses
by measuring just two $M_{T2}$ kinematic endpoints,
with arbitrary choices for the test mass $\tilde M_c$ and 
the upstream $P_T$. For concreteness, let us pick
some fixed $\tilde M'_c$ and $P'_T$,
form the corresponding $M_{T2}$ distribution (\ref{eq:mt2def})
and measure its endpoint $\tilde M'_p$, also making a note of 
the configuration $s'$:
\begin{equation}
\left\{ \tilde M'_c, P'_T\right\} \stackrel{measure}{\longrightarrow} \left\{\tilde M'_p, s'\right\}.
\label{meas1}
\end{equation} 
Now perform a second such measurement 
\begin{equation}
\left\{ \tilde M''_c, P''_T\right\} \stackrel{measure}{\longrightarrow} \left\{\tilde M''_p, s''\right\}.
\label{meas2}
\end{equation} 
By inverting (\ref{eq:Mp}), 
these two measurements allow the experimental determination of
\begin{equation}
p_{\ell T}(s'P'_T) = 
\frac{ \tilde M'^2_p-\tilde M'^2_c}{4\tilde M'^2_p}
\left( \sqrt{4\tilde M'^2_p+(s'P'_T)^2}-s'P'_T\right)
\label{eq:ptsprime}
\end{equation} 
and similarly for $p_{\ell T}(s''P''_T)$. Now taking the ratio
\begin{equation}
r \equiv \frac{p_{\ell T}(s'P'_T)}{p_{\ell T}(s''P''_T)}
= \frac{\sqrt{4M_p^2+(s'P'_T)^2}-s'P'_T}{\sqrt{4M_p^2+(s''P''_T)^2}-s''P''_T}\, ,
\label{eq:rdef}
\end{equation}
where in the second step we used eq.~(\ref{eq:ptlep}), 
we can solve (\ref{eq:rdef}) for the {\em true} 
parent mass $M_p$ in terms of measured quantities:
\begin{equation}
M_p = \left\{ \frac{-rs'P'_{T} s''P''_{T}}{(1-r^2)^2}
          \left(r-\frac{s'P'_{T}}{s''P''_{T}}\right)
          \left(r-\frac{s''P''_{T}}{s'P'_{T}}\right)\right\}^\frac{1}{2},
\label{Mp2method}
\end{equation}
and then find the {\em true} child mass $M_c$ from (\ref{eq:ptlep}) as
\begin{equation}
M_c =
M_p \left[ 1 - \left(1-\frac{\tilde{M}'^2_c}{\tilde{M}'^2_p}\right)
\frac{\sqrt{4 \tilde M'^2_p+(s' P'_T)^2} -  s'P'_T}
{\sqrt{4 M_p^2+(s'P'_T)^2} - s' P'_T} \right]^{\frac{1}{2}}
\label{Mc2method}
\end{equation}
with $M_p$ already given by (\ref{Mp2method}). 
Note than in this method, the values of
$\tilde M'_c$, $\tilde M''_c$, $P'_T$ and $P''_T$
can be chosen at will, allowing
for repeated measurements of $M_p$ and $M_c$.

\begin{figure}[t]
\includegraphics[width=7.5cm]{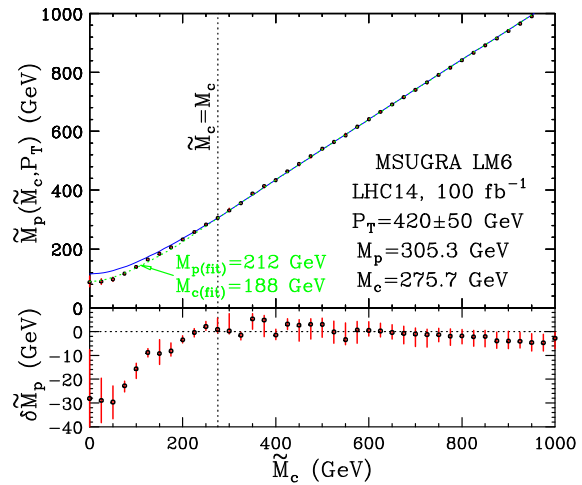}
\caption{\label{fig:kink} $M_{T2}^{max}$ versus
the test mass $\tilde M_c$, as obtained in
our simulations (data points) from a sample with
$P_T=420\pm50$ GeV, or theoretically
from eq.~(\ref{eq:MT2maxLR}) (blue solid line),
as well as their difference (lower panel).
}
\end{figure}

{\bf Method III.}
The third and final method for extracting the two masses $M_p$ and $M_c$
will make use of the celebrated ``kink'' in the $M_{T2}$ endpoint 
function (\ref{eq:MT2maxLR}) \cite{kink}.
Since $\tilde M_p (\tilde M_c, +P_T)$
and $\tilde M_p (\tilde M_c, -P_T)$ have
different slopes at the crossover point $\tilde M_c= M_c$,
the function $\tilde M_p (\tilde M_c, P_T)$
has a slope discontinuity precisely at the correct value 
$M_c$ of the child mass, providing an alternative measurement of 
the absolute mass scale \cite{kink}. 
The procedure is illustrated in 
Fig.~\ref{fig:kink} for the LM6 study point of Table~\ref{tab:mass}.
The blue solid line shows the theoretically expected
shape from eq.~(\ref{eq:MT2maxLR}), for $P_T=420$ GeV,
which is roughly the mean of the $P_T$ distribution at 
point LM6. In the LM6 case the kink is very mild, only $3.3^\circ$ \cite{Burns:2008va}. 

In order to test the precision of the three methods, we perform 
event simulations using the PYTHIA event generator \cite{Sjostrand:2006za}
and PGS detector simulation \cite{PGS}.
We consider the LHC at its nominal energy of 14 TeV and 
$100\ {\rm fb}^{-1}$ of data.
To ensure discovery, we use standard CMS cuts as follows 
\cite{Ball:2007zza,Pakhotin:2006wx}: exactly
two isolated leptons with $p_T>10$ GeV, 
at least three jets with $p_T>(175,130,55)$ GeV,  
$\met > 200$ GeV and a veto on tau jets.
With those cuts, in the dimuon channel alone, the 
remaining SM background cross-section is rather 
negligible (0.15 fb), while
the SUSY signal is 14 fb, already leading to a
$22\sigma$ discovery with just $10\ {\rm fb}^{-1}$ 
of data \cite{Ball:2007zza,Pakhotin:2006wx}.
In order to compare to the theoretical result in
Fig.~\ref{fig:kink}, we select a $\pm50$ GeV 
$P_T$ bin around $P_T=420$ GeV and construct 
a series of $M_{T2}$ distributions, for different 
input values of $\tilde M_c$. For each case, 
we include all SM and SUSY combinatorial backgrounds, 
and extract the $M_{T2}^{max}$ endpoint by a linear 
unbinned maximum likelihood fit, obtaining
the data points shown in Fig.~\ref{fig:kink}. We see that
the $M_{T2}$ endpoint can be determined rather well
($\delta \tilde M_p\lsim 3$ GeV),
but {\em only} on the right branch $\tilde M_c\ge M_c$.
\begin{figure}[t!]
\includegraphics[width=7.5cm]{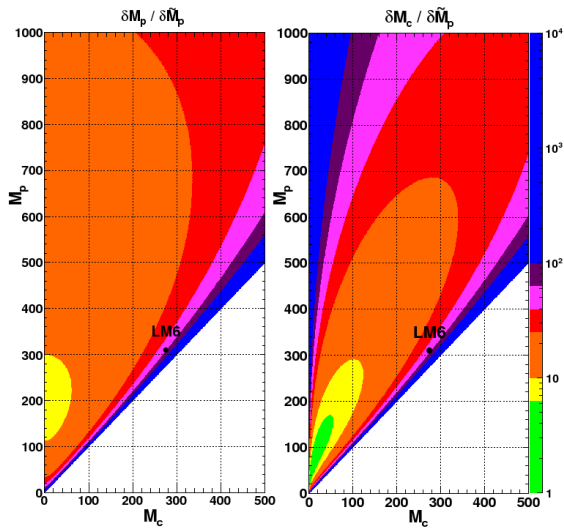}
\caption{\label{fig:err} 
Scaling factors relating the error $\delta \tilde M_p$
in the extraction of the $M_{T2}$ endpoint to the
resulting uncertainties $\delta M_p$ and $\delta M_c$ 
on the parent and child masses calculated from (\ref{Mp2method})
and (\ref{Mc2method}), as a function of the true
input masses $M_c$ and $M_p$.
}
\end{figure}
In contrast, the $M_{T2}$ endpoints on the
left branch $\tilde M_c\le M_c$ are considerably
underestimated, washing out the expected kink.
There are two separate 
reasons 
behind this effect. Recall that 
the $M_{T2}$ endpoint on the left branch 
is obtained in the configuration $s=+1$
of Fig.~\ref{fig:mom}, 
which requires 
the lepton to be emitted in the backward direction.
As a result, the parent boost favors configurations
with $s\simeq -1$ over $s\simeq +1$. 
Another consequence is that leptons with 
$s\simeq +1$ are softer and
more easily rejected by the offline $p_T$ cuts.
We conclude that $M_{T2}^{max}$ measurements 
on the left branch are in general not very reliable,
and tend to jeopardize the traditional kink method.
For example, using Method III to fit the data in Fig.~\ref{fig:kink}
(green dotted line), we find best fit values of 
only $M_{p(fit)}=212$ GeV and $M_{c(fit)}=188$ GeV.
Method I has a similar problem, since
$p_{\ell T}(+P_T)$ is measured from events 
in the $s=+1$ configuration. 
Using the $\tilde M_p$ measurements 
from Fig.~\ref{fig:kink} at
$\tilde M_c=0$ and $\tilde M_c=1$ TeV, we find from 
eq.~(\ref{eq:ptsprime}) that
$p_{\ell T}(+420\ {\rm GeV})=8.8$ GeV
and 
$p_{\ell T}(-420\ {\rm GeV})=50.6$ GeV
(compare to the nominal values of 
14.8 GeV and 53.6 GeV, correspondingly).
The resulting mass determination 
via eqs.~(\ref{Mptrue},\ref{Mctrue})
is $M_{p(fit)}=212$ GeV and $M_{c(fit)}=190$ GeV.
We see that in both Method I and Method III,
the masses are underestimated due to the 
systematic underestimation of the 
left $M_{T2}^{max}$ branch in Fig.~\ref{fig:kink}.
It is therefore of great interest to have an 
alternative method, which relies
on the right $M_{T2}^{max}$ branch alone.

This is where 
the available freedom in Method II comes into play,
since {\em both} test masses $\tilde M'_c$ and $\tilde M''_c$
can be chosen on the right branch. Taking 
$P'_T=350\pm50$ GeV and $P''_T=500\pm 50$ GeV
and repeating our earlier analysis, we find that
$\delta \tilde M_p$ on the right branch 
is still on the order of 3 GeV, as in Fig.~\ref{fig:kink}.
The resulting error $\delta M_p$ ($\delta M_c$)
on the measured parent (child) mass can be easily
propagated from eqs.~(\ref{Mp2method},\ref{Mc2method}).
The two ratios $\delta M_p/\delta\tilde M_p$
and $\delta M_c/\delta\tilde M_p$
are shown in Fig.~\ref{fig:err}, where 
for concreteness we have taken 
$\tilde M'_c=\tilde M''_c=1000$ GeV.
Fig.~\ref{fig:err} reveals that the 
LM6 input values of $M_c$ and $M_p$ are 
rather unlucky, since the error 
$\delta \tilde M_p$ on the $M_{T2}$ 
endpoint is then amplified by a factor of almost 70.
However, if $M_c$ and $M_p$ happened to be
different, with the rest of the spectrum the same, the
precision quickly improves.
For example, with $\delta \tilde M_p=\pm 3$ GeV,
the masses can be determined to within
$\pm 30$ GeV ($\pm 75$ GeV) within the yellow (orange)
region.
One should keep in mind that the dominant uncertainty on
$\delta \tilde M_p$ is due to the SUSY combinatorial background.
We have verified that in the absence of such combinatorial 
background, $\delta \tilde M_p\lsim 1$ GeV and the
typical precision on $M_p$ and $M_c$ 
from Fig.~\ref{fig:err} is then at the level of 10\%.



In conclusion, we considered the 
inclusive same-sign dilepton channel in SUSY, which so far has only been 
used for discovery, but not for mass measurements.
We demonstrated that it allows a separate determination of
the chargino and sneutrino masses.
We discussed three different methods, which rely exclusively
on the well measured lepton momenta.
The methods are completely general and inclusive, and can be 
applied to other SUSY topologies
and to non-SUSY scenarios like UED \cite{ued}. 

{\bf Acknowledgements.}
The work of KM and MP is supported by a US DoE grant DE-FG02-97ER41029.


\end{document}